\documentclass[twocolumn,secnumarabic,amssymb, nobibnotes, aps, prd]{revtex4}
\usepackage{graphicx}
\usepackage{color}

\newcommand{\beq}{\begin{equation}}
\newcommand{\eeq}{\end{equation}}
\newcommand{\beqa}{\begin{eqnarray}}
\newcommand{\eeqa}{\end{eqnarray}}

\begin{document}          
\title{Quark-gluon plasma paradox}
\author{D.~Mi\'{s}kowiec}
\email{d.miskowiec@gsi.de}
\affiliation{Gesellschaft f\"ur Schwerionenforschung mbH, 
Planckstr. 1, 64291 Darmstadt}
\begin{abstract}
Based on simple physics arguments it is shown that the concept of 
quark-gluon plasma, a state of matter consisting of uncorrelated quarks, 
antiquarks, and gluons, has a fundamental problem. 
\end{abstract}

\maketitle

\section{Introduction}
The existence of quark-gluon plasma (QGP), a state of matter in which 
quarks are free to move in space, was postulated by Cabibbo and Parisi 
in \cite{cabibbo} in response to the concept of limiting temperature 
of Hagedorn \cite{hagedorn1,hagedorn2}. 
A transition from hadronic matter to QGP is supposed to occur at an 
energy density of about 1~GeV/fm$^3$ which can be reached either by heating 
or by compressing or both. 
Intensive search for QGP in collisions of lead and gold nuclei at 
energies of up to $\sqrt{s_{NN}}=$ 17.2~GeV at the CERN SPS yielded 
``compelling evidence for the existence of a new state of quark-gluon matter 
(...) in which quarks are liberated to roam freely'' \cite{cern}. 
The subsequent Au+Au collision experiments at RHIC, 
albeit with much higher energies of $\sqrt{s_{NN}}\leq $ 200~GeV, 
resulted in a somewhat weaker statement reporting only a 
``new form of nuclear matter'' with an ``energy density and temperature 
clearly exceeding the critical values predicted by QCD calculations'' 
\cite{rhic}. 
In the same report it is stressed that the observed medium behaves like 
a strongly coupled fluid rather than the expected gas of free uncorrelated 
quarks. 
The common expectation is that the latter can still exist at 
energies several times higher than the critical energy density \cite{kapusta}. 
This is supported by lattice calculations which show that with increasing 
temperature the system (slowly) approaches the ideal gas limit 
\cite{karsch-qm2006} 
albeit deviations from ideal gas still occur at 
temperatures as high as (2-3) $T_c$ \cite{karsch-hard2006}. 
The evidence and the possible reasons for these deviations were extensively 
discussed by a pioneer of the field in \cite{shuryak}. 

In this letter I use simple physics arguments to show that the concept of 
QGP, a state of matter with liberated quarks, at any temperature has a 
fundamental problem. 
The problem, 
which does not manifest itself during creation of QGP but only 
during the transition back to hadrons, 
consists in the fact that 
simultaneous hadronization in regions separated by space-like intervals 
must in some cases lead to single quarks left at the borders between 
hadronization domains 
because there is no way to synchronize this process without violating 
causality. 
The problem is exposed in detail in Section~2 by means of a 
gedanken experiment. 
In Section~3 I will discuss possible solutions of the paradox. 

\section{Demonstration of the problem}
I start from the assumption that the QGP, large (comparing to a nucleon) 
volume filled with uncorrelated quarks, antiquarks, and gluons, can exist. 
I will then use a certain amount of QGP to perform a gedanken experiment 
during which I only do things which are not forbidden by physics laws. 
The final state after the experiment, nevertheless, will be one with 
isolated quarks separated by a macroscopic distance which is not allowed by 
QCD. Here are the steps of the procedure.

i) I create one cubic mm of QGP with a temperature well above the critical 
temperature and a total net baryon number $\mu$=0. 
I stretch it to dimensions of 10 fm x 10 fm x 1000 light years, keeping 
the density constant. 
I connect both ends such as to form a ring. 

ii) I break the QGP ring at one point by allowing the QGP to expand and cool 
such that the hadronization starts there.
The phase boundary propagates along the ring in both directions with the 
velocity of, say, 0.05~c. 
For the problem under discussion it does not matter whether the propagation 
of the phase boundary is spontaneous (rarefaction wave, moving with the speed 
of sound) or imposed from outside (removing the bonds). 
The hadronization proceeds until the last chunk of QGP, on the opposite side 
of the ring, turns into hadrons. 

\begin{figure}[b]
\includegraphics[width=4.1cm,height=3cm]{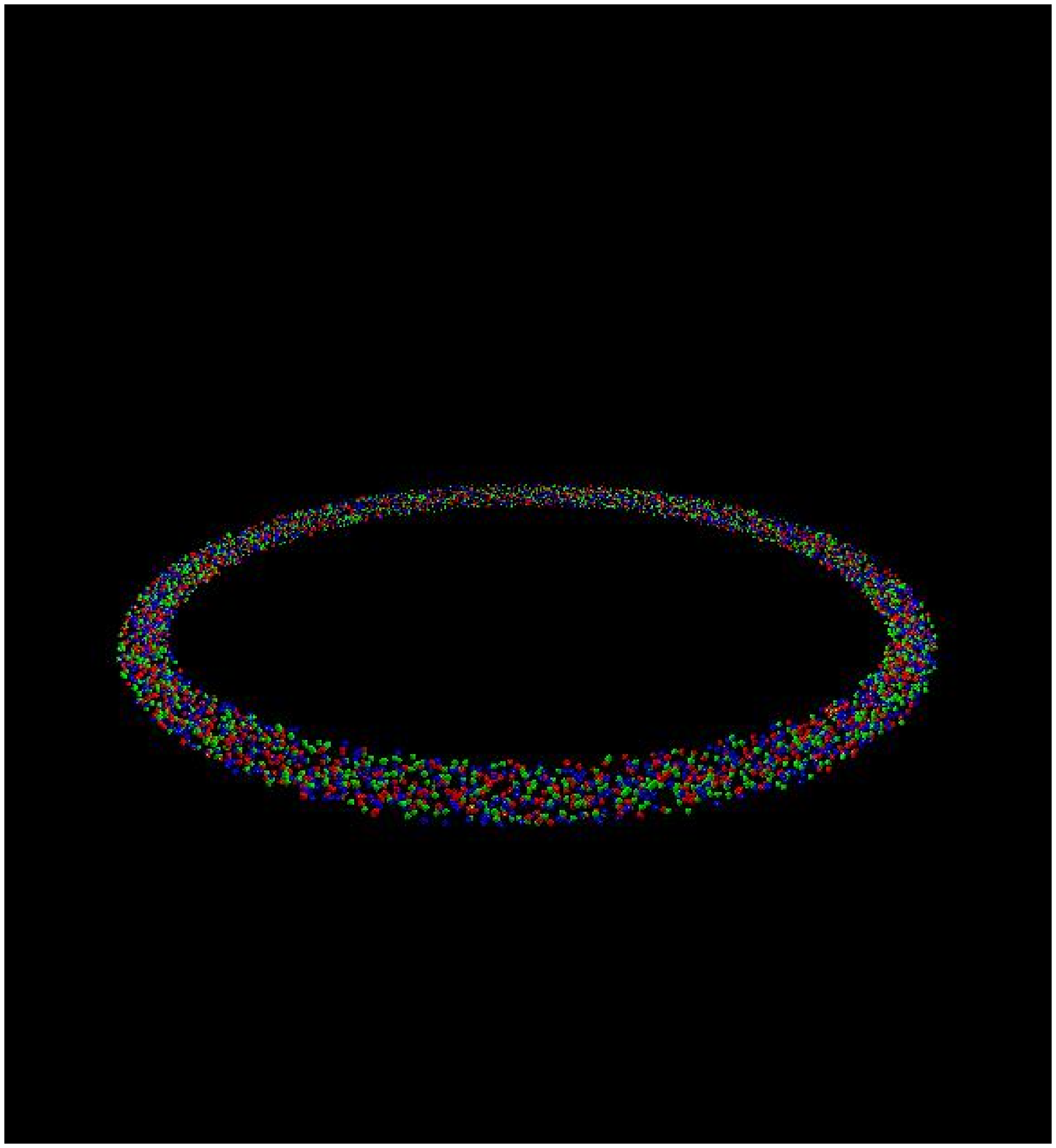} 
\includegraphics[width=4.1cm,height=3cm]{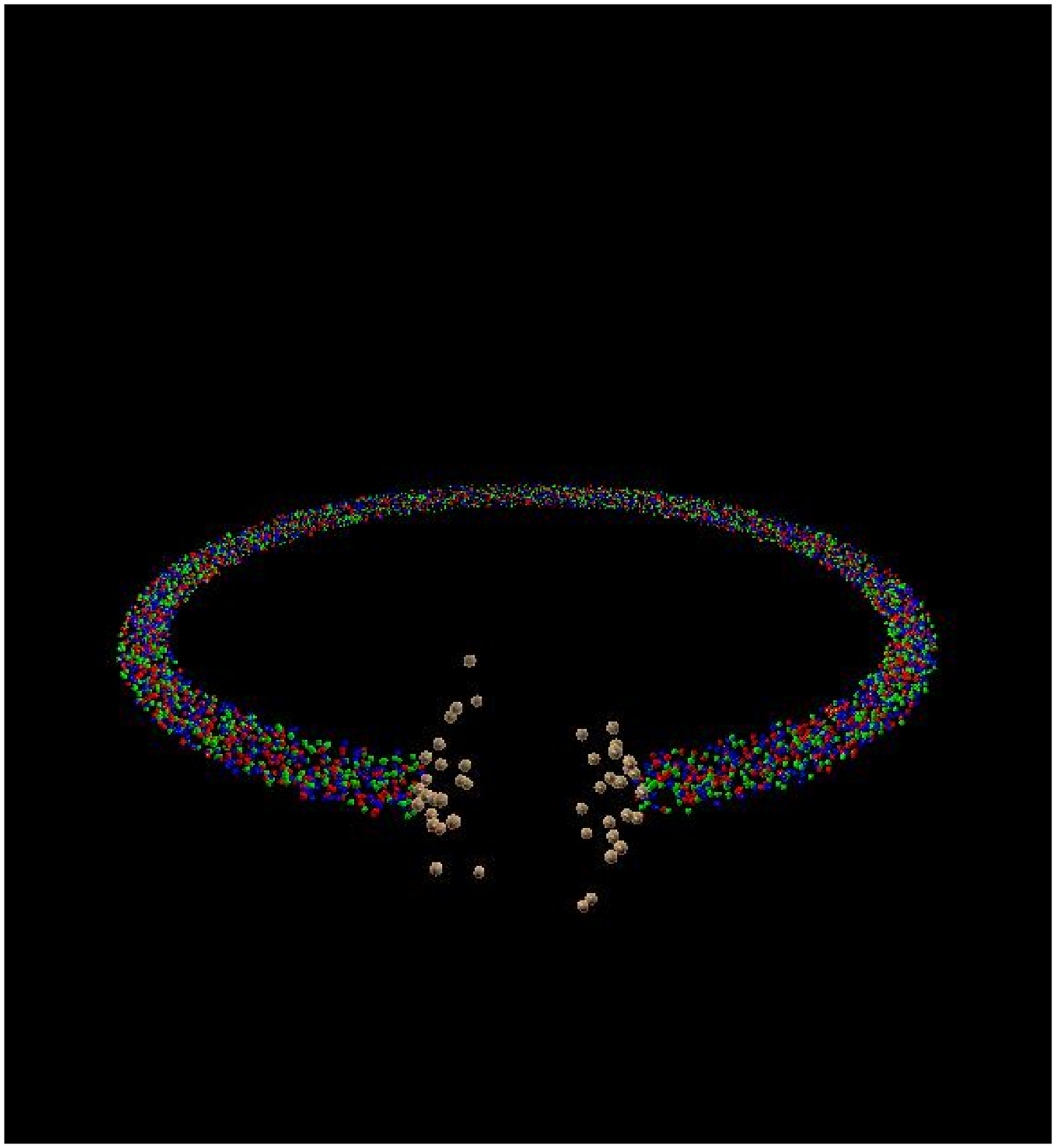} \\ \vspace*{0.5mm}
\includegraphics[width=4.1cm,height=3cm]{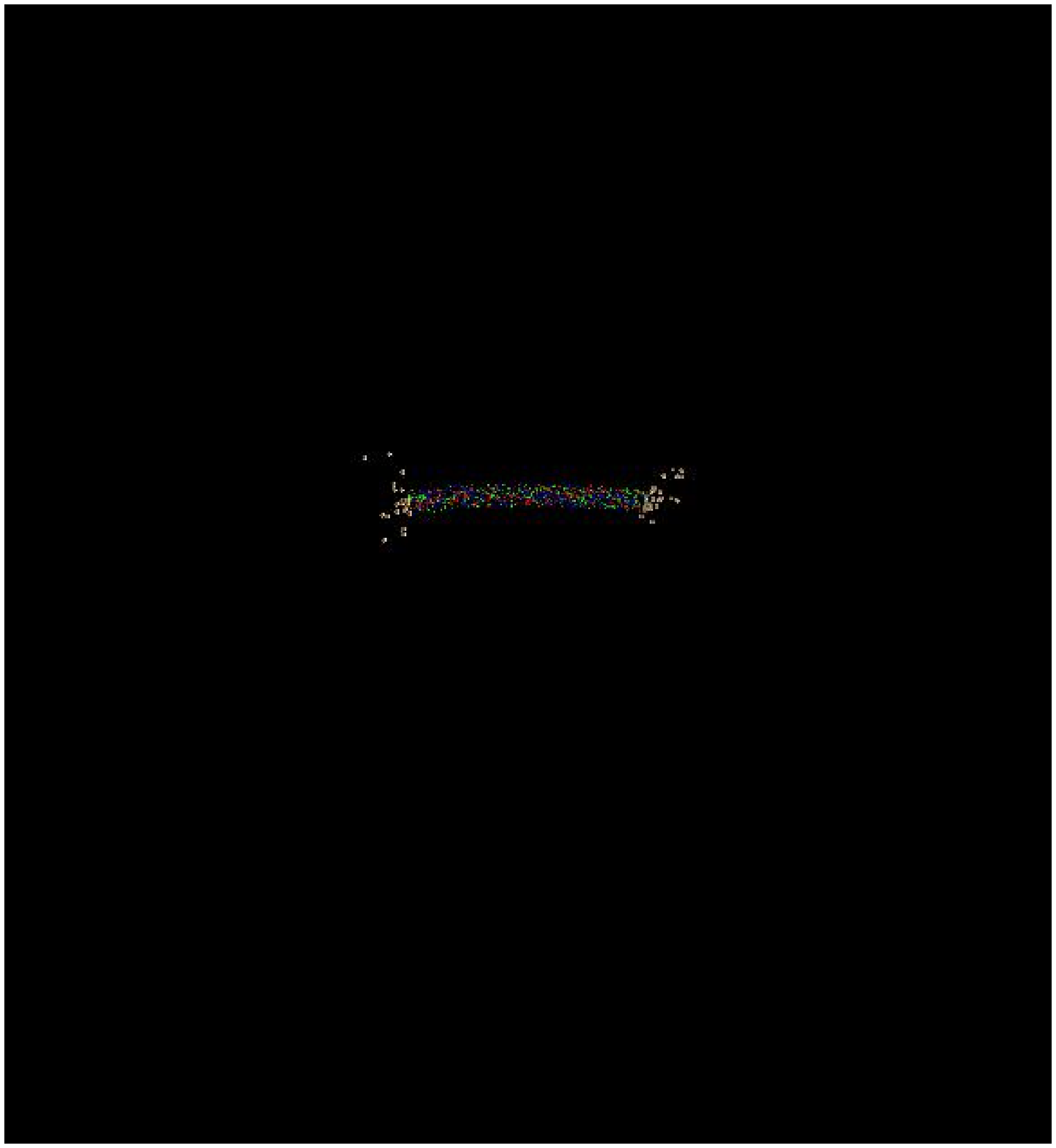} 
\includegraphics[width=4.1cm,height=3cm]{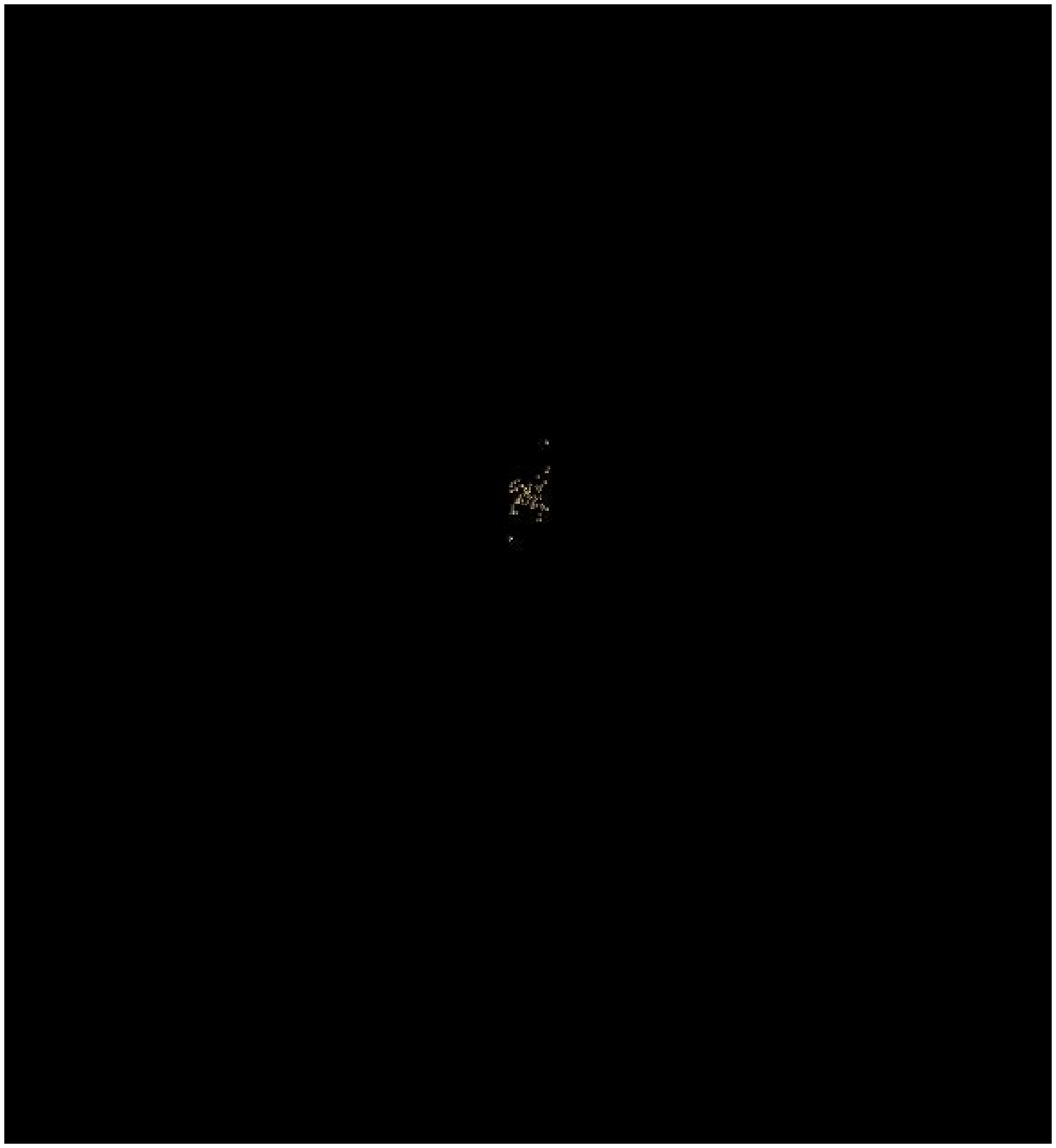}
\caption{Converting QGP into hadrons, scenario 1. 
The hadronization starts at one point of the ring and propagates along 
the ring in both directions. (Colors online.)}
\label{fig:ring1}
\end{figure}

I can repeat this gedanken experiment, pictorially represented in 
Fig.~\ref{fig:ring1}, many times. 
The hadronization is always successful in the sense that all quarks 
in the system are turned into hadrons. 
Now, however, I introduce a little modification in the second step:

ii') 
As before, 
I break the QGP ring at one point by allowing the QGP to expand and cool such 
that the hadronization starts there.
At the same time \footnote{
To talk about ``time'' I need to specify a reference frame. 
Let us pick the frame with the origin located in the middle of the ring
and in which the total momentum of the QGP is zero.}, 
however, my assistant does the same at the opposite end of the ring. 
This has no immediate influence at what is happening at my end of the ring
because the two points are separated by light years. 

\begin{figure}[b]
\includegraphics[width=4.1cm,height=3cm]{kuku00.eps} 
\includegraphics[width=4.1cm,height=3cm]{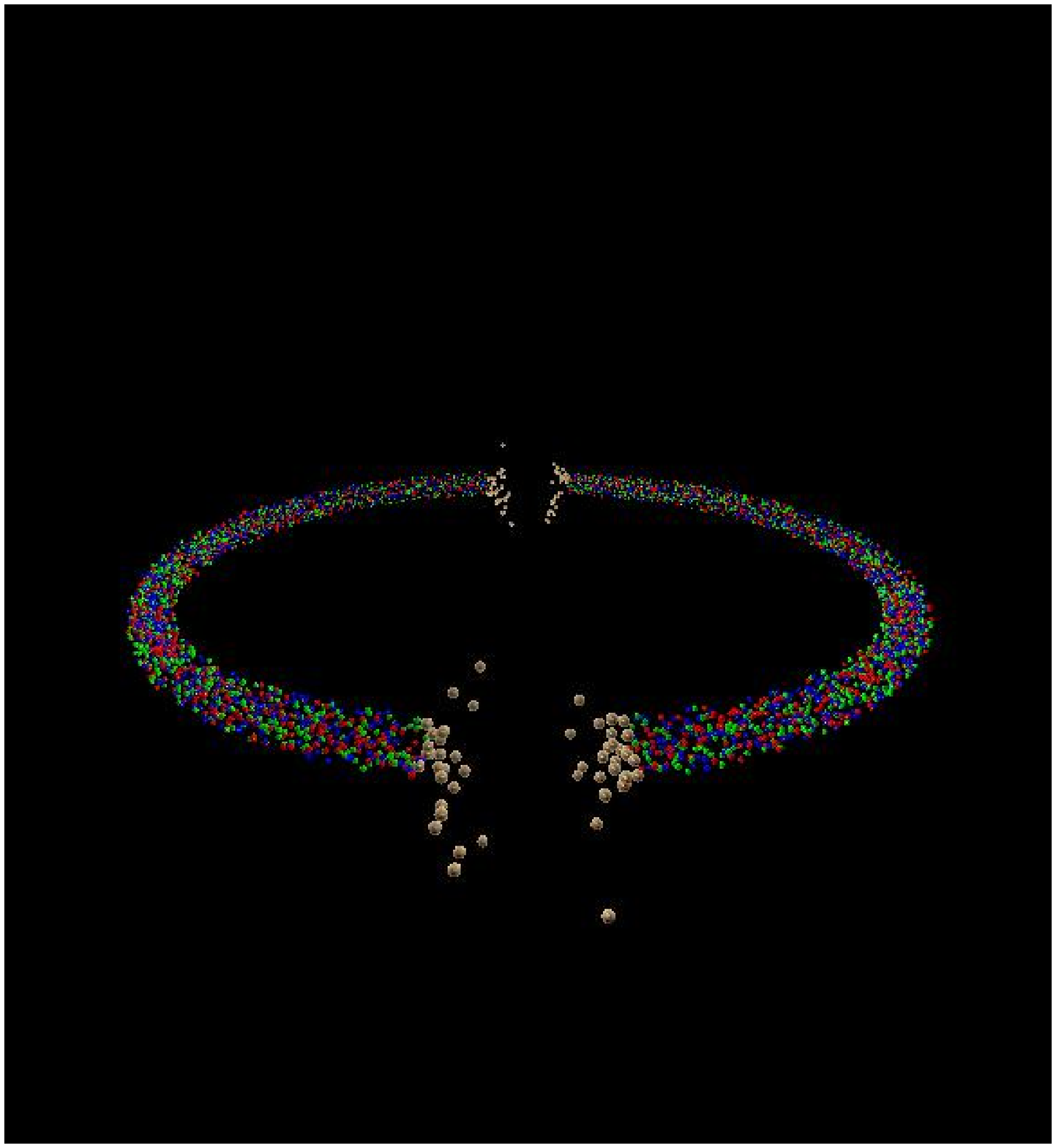} \\ \vspace*{0.5mm}
\includegraphics[width=4.1cm,height=3cm]{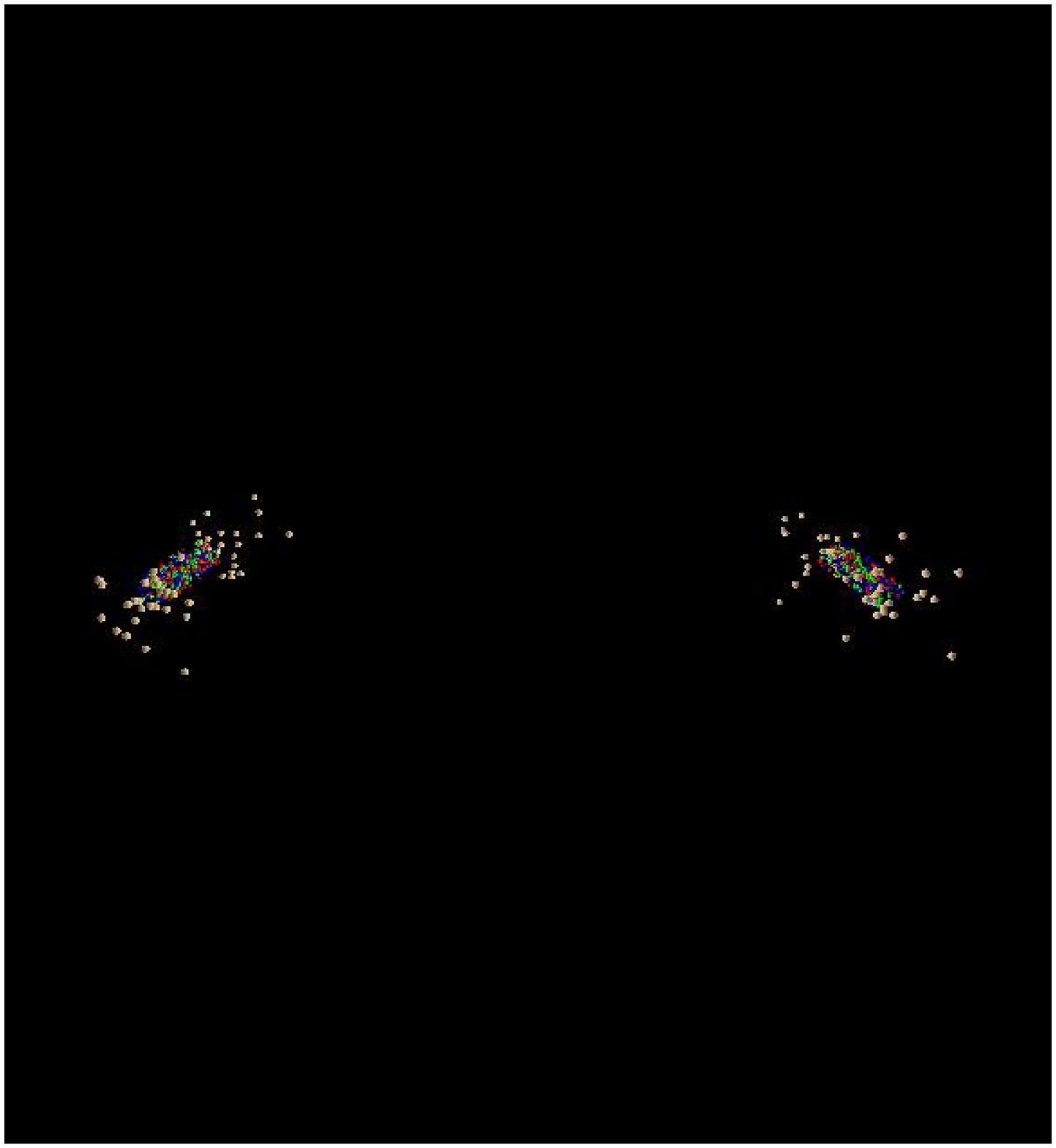} 
\includegraphics[width=4.1cm,height=3cm]{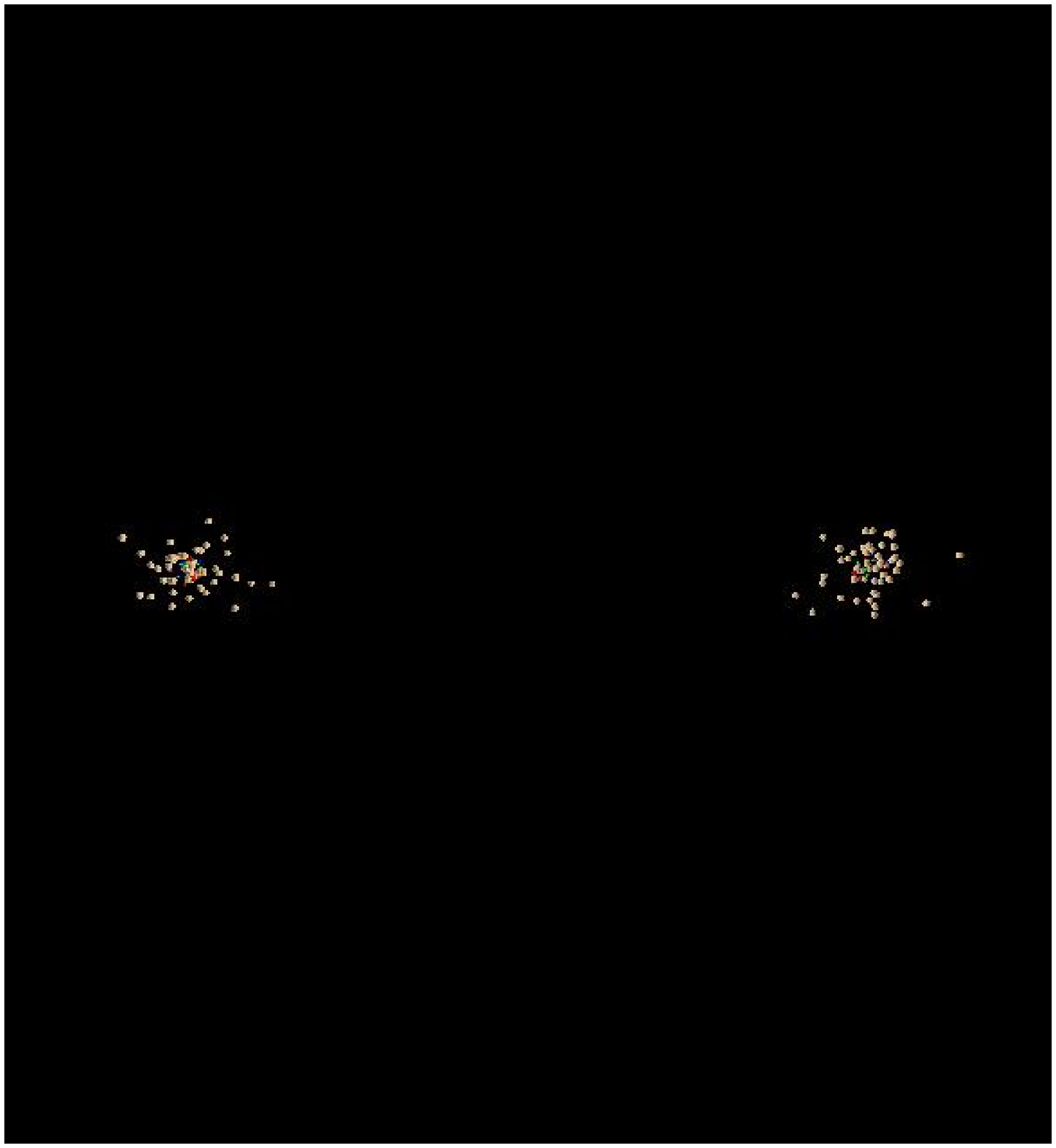}
\caption{Converting QGP into hadrons, scenario 2. The hadronization starts 
at two opposite points of the ring, separated by 300 light years, and 
propagates from each starting point in both directions. Whether the two 
created QGP blobs have integer or fractional net baryon numbers depends on 
the choice of the two starting points so this information is not available 
earlier than a couple of hundreds of years after the start of hadronization. 
By that time the QGP blobs are separated by such a distance that a string 
between them would require too much energy. (Colors online.)}
\label{fig:ring2}
\end{figure}

Now I have two separate blobs of QGP. The four phase boundaries propagate 
until two small chunks of QGP remain (Fig.~\ref{fig:ring2}).

Obviously, there is a 33\% chance that these two chunks have integer net 
baryon numbers. With the remaining probability of 67\% they have fractional 
ones. 
So, if I repeat our second experiment many times, 
sooner or later I will end up with two objects with fractional baryon 
numbers, separated by light years. This state is not allowed by QCD. 

We started from an allowed state, we never did anything forbidden by 
physics laws, and we ended up 
\footnote{More precisely, we have a vanishing chance of avoiding the forbidden 
state if we repeat the experiment many times.} 
in a state which is forbidden. 
In the next section I will discuss possible resolutions of the paradox. 
Before doing this, several remarks are in order 
regarding the technical aspects of the presented gedanken experiment. 

First, 
while I was considering here the baryon numbers of the outcoming 
particles and requiring it to be integer, 
alternatively, one could monitor the color of the final particles, 
and require them to be white. 
In both cases the reasoning is equivalent and leads to the same conclusions. 
For technical reasons I decided to base the argument on baryon number 
and not on color -- except in Figs.~\ref{fig:zoo} and \ref{fig:zoo2} where 
color is better to explain the point. 

Second, 
the amount of QGP used in the described gedanken experiment is much 
higher than the one we are familiar with, i.e. the one expected in a 
relativistic heavy ion collision, and the ring-like shape is something 
one would not expect to be very frequent in nature. 
On the other hand, a simple calculation using the mass of the observable 
universe indicates that the amount of QGP during Big Bang was much higher 
than the one considered here. 
What concerns the ring shape, while a ring is best to illustrate the problem, 
the problem remains the same even if one squeezes the ring such that the two 
hadronization starting points get close to each other. 
In this case the QGP blob resembles in shape the elongated fireball created 
in a heavy ion collision, with the hadronization starting in the middle 
(also quite possible in a heavy ion collision). 
The information does not need to propagate from the other side of the ring 
but only across some 5-10~fm, so the situation quantitatively 
is much less dramatic. Qualitatively, however, the problem is the same - 
whether or not given quarks are allowed to form a hadron depends not only on 
themselves and on their direct neighborhood but also on remote parts 
of the QGP volume. One could argue that this is not a problem on the 
scale of 5-10 fm. However, the QGP blob created in the Early 
Universe had the same problem if the hadronization, caused by the 
expansion, was happening in the whole volume. 
Converting quarks into hadrons in the whole volume at the same time can 
be compared to trying to reach a homogeneous magnetization in 
a bulk ferromagnetic by lowering the temperature. 

Third, 
since the plasma temperature does not enter explicitely the problem persists 
for all temperatures above the critical temperature. 
In particular, one cannot argue that the paradox is restricted to the 
cases with temperatures close to the critical temperature. 

Fourth,
one could try to dismiss the depicted experiment by arguing that splitting QGP 
in two parts is like splitting conducting material in electrodynamics. 
There, 
any residual charge imbalance is removed by the current flowing through 
the last point of contact between the two halves. 
However, this is true only if the detaching proceeds slowly. 
If e.g. two metals with different work functions are detached quickly 
non-zero net charges remains on the two parts. 
For electric charge, unlike baryon number, this is not a problem. 

\section{Possible solutions}
\label{section:solution}

It is worth considering whether the described problem could be just  
another case of the famous EPR paradox. 
The QGP ring is an entangled state of quarks and gluons. 
When hadronization starts at one point the wave function collapses 
and from now on every point of the ring ``knows'' whether starting 
hadronization there is allowed or not. 
However, in this case one could use the ring for superluminal transfer 
of information. Indeed, if upon starting hadronization I observe 
a string then this means that my assistant did start 
hadronization, even if he did it only one second before 
me and even if he is many light years away from me. 
It is commonly believed, however, that the entangled states and the EPR 
paradox do not offer a way of propagating information at superluminal 
speed \cite{tiwari}. 

The second way out would be to assume that the QGP properties are such that 
a plasma blob cannot have a hole, and the hadronization can only 
happen at the surface. 
Volume hadronization, 
e.g. caused by  density dropping uniformly in the entire volume during 
Hubble-like expansion, would be forbidden. 
The QGP would be resistant against attempts of pulling 
apart pieces of it, i.e. it would behave like a liquid with infinite surface 
tension. 
For heavy ion collisions it would mean that the hadronization starts 
at both ends of the elongated fireball, in spite of the fact that 
the particles there have the largest Lorentz gamma factor. 
For the Early Universe the consequences are much more dramatic. 
Since the phase boundary cannot proceed faster than the speed of sound, 
and certainly not faster than the speed of light in vacuum, 
and since the observable universe in the QGP state 
had dimensions comparable to the size of the Solar System, its hadronization 
must have taken minutes. 
Since the entire universe may be much larger than the observable universe 
(maybe infinite) the actual time needed might be even longer. 

\begin{figure}[ht]
\vspace{5mm}
\includegraphics[height=4cm]{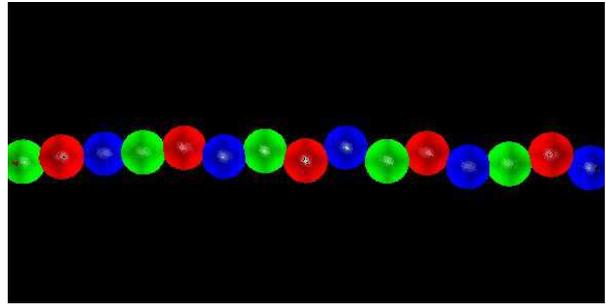}
\caption{A very thin ring of QGP, zoomed. The balls represent quarks. 
In this figure antiquarks and gluons were left out so what is represented 
is a cold and dense QGP rather that the hot and symmetric QGP discussed 
throughout the note. For the latter the argument would be the same but the 
corresponding figure would be more difficult to draw. (Colors online.)}
\label{fig:zoo}
\end{figure}

\begin{figure}[ht]
\vspace{5mm}
\includegraphics[height=4cm]{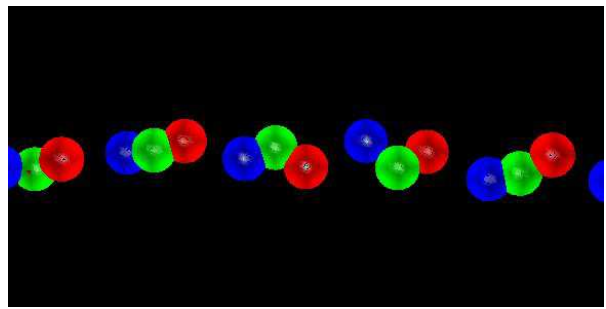}
\caption{A very thin ring of QGP, zoomed. The balls represent quarks. 
The quarks are grouped into white clusters with integer baryon number. 
(Colors online.)}
\label{fig:zoo2}
\end{figure}

The third possibility is that local correlations between quarks make some 
cutting surfaces more probable than the others when it comes to cutting 
the ring and starting the hadronization. 
Obviously, in absence of such correlations the QGP ring basically looks 
like in Fig.~\ref{fig:zoo} and no preferred breaking points can be 
recognized. 
If, however, some kind of interactions lead to clustering of quarks and gluons 
into (white) objects of integer baryon numbers like in Fig.~\ref{fig:zoo2} 
then starting hadronization from several points of the ring at the same time 
will not lead to any problem. 
However, this kind of matter would be hadron resonance matter rather than 
the QGP. 
The degrees of freedom would not be the ones of quarks and gluons, 
expected from a genuine quark-gluon plasma. 
Arguing that the plasma may look like in Fig.~\ref{fig:zoo} at high 
temperatures and like in Fig.~\ref{fig:zoo2} close to the phase transition 
does not resolve the paradox because the transition from uncorrelated 
quarks to clusters again has to take a time comparable 
to the size of the QGP volume divided by the speed of light. 
The much shorter time scales of ``whitening of the QGP'', obtained in 
\cite{mrowczynski}, were based on statistical considerations 
in which the problem discussed in this letter can not show up. 

\section{Summary}
I demonstrated that the concept of QGP, state of matter with 
uncorrelated quarks, antiquarks, and gluons, leads to isolated 
objects with fractional baryon numbers, 
unless superluminal signalling is allowed, or, by some 
mechanism, the hadronization is restricted to the surface of 
the QGP volume, meaning that e.g. the hadronization in the Early 
Universe took at least minutes rather than a couple of microseconds. 
The third, obvious, way of avoiding the paradox is to declare the  
uncorrelated QGP as non-existent, and to replace it by a state consisting 
of quark clusters with integer baryon numbers (resonance matter). 
Both the surface-hadronization and the resonance matter options 
result in a liquid- rather than a gas-like structure of the matter. 
This agrees with the hydrodynamical character of the matter created in 
nuclear collisions at RHIC and, at the same time, indicates that this 
character will be preserved at higher temperatures. 

I gratefully acknowledge useful discussions with Peter Braun-Munzinger, 
Pawe\l{} Danielewicz, Staszek Mr\'owczy\'nski, Sergei Voloshin, and, 
especially, Uli Heinz. 
At the same time I would like to stress that they carry no responsibility 
for the views expressed in this letter. 
I would also like to thank Bengt Friman and Anton Andronic for reading 
the manuscript and helping to spot its many deficiencies.

\end{document}